\documentclass[twocolumn,letterpaper,amsmath,amssymb,floatfix,prl,superscriptaddress]{revtex4} 

\usepackage{graphicx}
\usepackage{dcolumn}  % Align table columns on decimal point
\usepackage{bm}  % bold math
\usepackage{epsfig}

\begin{document}
\setlength\arraycolsep{2pt}

\title{Drag forces in classical  fields}

\author{Vincent D\'emery}
\affiliation{Universit\'e de Toulouse, UPS, Laboratoire de Physique Th\'eorique (IRSAMC),  CNRS UMR5152, F-31062 Toulouse, France}

\author{David S. Dean}
\affiliation{Universit\'e de Toulouse, UPS, Laboratoire de Physique Th\'eorique (IRSAMC),  CNRS UMR5152, F-31062 Toulouse, France}

\begin{abstract}
Inclusions, or defects, moving at constant velocity through free classical  fields are shown to be subject to a drag force which depends on the field dynamics and the coupling of the inclusion to the field.  The  results are used to predict the drag exerted on inclusions, such as proteins, in lipid membranes  due to their interaction with height and composition fluctuations. The force, measured in Monte Carlo simulations, on a point like magnetic field moving through an Ising ferromagnet is also well explained by these results.
\end{abstract}

\maketitle
In quantum field theory  forces between particles are induced via their coupling to a  quantum field \cite{wein}. The same phenomena  arises  for fields driven by thermal  fluctuations, for example interactions between inclusions in fluctuating 
membranes \cite{gou1993}. Similarly the Casimir force, both quantum and thermal, arises due to the imposition of boundary conditions on  quantum or thermal fields \cite{kar1999}. Casimir
discovered his force in the quantum context but Fisher and de Gennes \cite{fi1978} showed that a classical version of this effect should be expected for fluctuating thermal fields, such as those for the order parameter of systems near a critical point. This critical Casimir effect has only recently been measured \cite{he2008} and the technical progress involved in this experiment may well open up the possibility of exploring other aspects of the critical Casimir effect, notably dynamical phenomena. 
In all of the above,  the effect of the field is manifested by the interaction induced between two or more particles or surfaces in the field. However the presence of the field can also be seen
by looking at the force exerted on a particle when it is not at rest.  For instance,  electrons moving in materials induce a local polarization  known as the polaron 
\cite{pol} which modifies their dynamics. A frictional Casimir force is also induced by the uniform motion of a conductor in a volume of blackbody radiation which is in equilibrium in the rest frame of a cavity containing the radiation \cite{mkr2003}. 

\begin{figure}
\epsfxsize=1\hsize \epsfbox{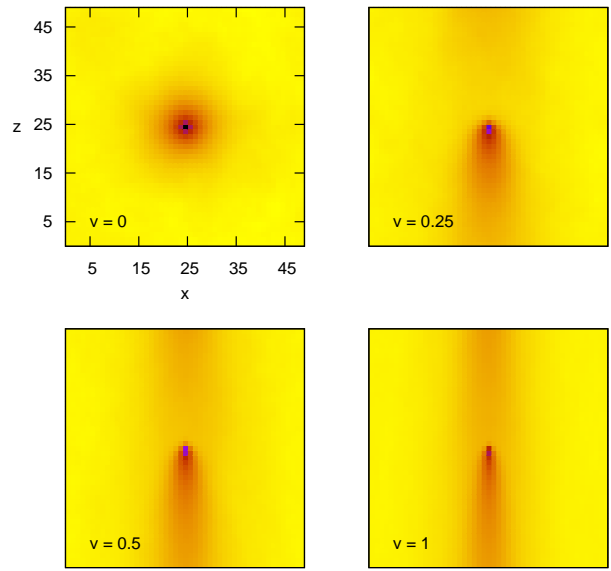}
\caption{Contour plot of the magnetization profile (polaron) for the 2d Ising model about a local magnetic
field at a single point moving with velocity $v$. 
 (high temperature phase: $\beta=1$, $J=0.4$, $h=6.66$). }
\label{fig1}
\end{figure}

In this letter we show that for classical  fields, in a laboratory rest frame, a drag force is present on inclusions linearly coupled to the field, and which move at constant velocity. The underlying physics is caused by  a polaron-like phenomena (see Fig. (\ref{fig1}))
which we generalize to a range of  statistical field models arising in the study of soft condensed matter systems. A key point in this analysis is that we  examine the effect of the dynamical models commonly used to  study soft condensed matter systems on the drag forces on inclusions.

As a concrete example of the class of problems we will 
address we  start by studying drag forces  in the Ising ferromagnet model on a $d$ dimensional
cubic lattice with Hamiltonian
\begin{equation}
H =-J\sum_{(i,j)} S_iS_j + hS_{i_0}.
\end{equation}
Here $J>0$ is a ferromagnetic coupling between nearest neighbor spins on a square lattice of
spacing $a$, and $h$ a magnetic field at the position of an inclusion $i_0$ which moves in
the direction $z$ with a velocity $v$,  so   $i_0(z) = {vt/a}$ (where we take the integer part). 
The system evolves in the following manner, the underlying unit of time is a Monte Carlo sweep
where $N$ (the system size) randomly chosen spins are examined and flipped or not  according to  the dynamical rules used: (i)  dynamics  not conserving the total magnetization 
- Glauber dynamics - a single spin is chosen and is flipped with probability
$p_f = 1/[1+\exp(\beta \Delta H)]$, where $\beta$ is the inverse temperature and $\Delta H$ the
energy change associated with the spin flip (ii) a form of Kawasaki dynamics conserving the total magnetization, here two spins of opposing sign are randomly selected and 
flipped with the probability $p_f$. After $\tau= a/v$ units of time the inclusion is moved one step in the $z$ direction.  The instantaneous force is defined as $f(t) = -\Delta E(t)/a=-h(S_{i_{0}+1}(t)-S_{i_{0}-1}(t))/2a$ (the energy change $\Delta E$ being computed from the symmetric discrete derivative on the lattice).

\begin{figure}
\epsfxsize=0.8\hsize \epsfbox{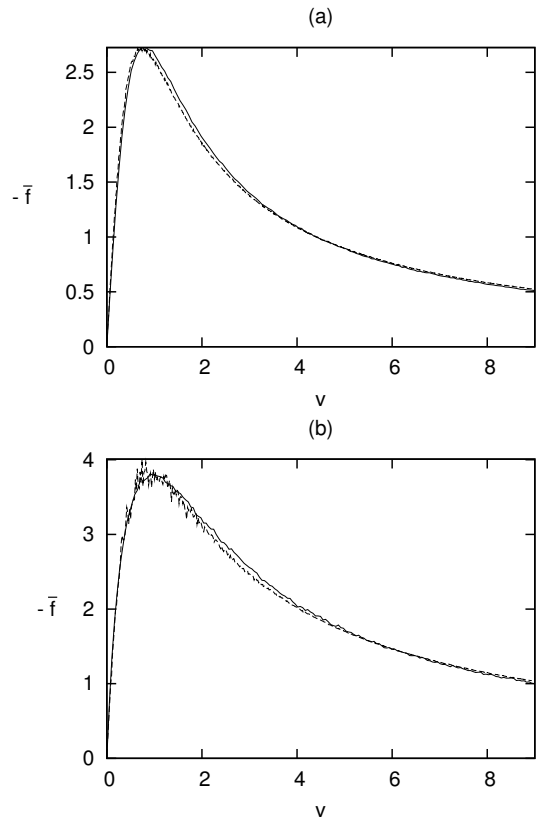}
\caption{Dashed lines: average drag force $\overline f$ in the 1d Ising model  as a function of $v$ for Glauber  (a) and Kawasaki dynamics (b). Solid lines are the fits of model A (a) and model B (b) dynamics
for the Gaussian ferromagnet.}
\label{fig2}
\end{figure}

Shown in  Figs. (\ref{fig2}a)
and (\ref{fig2}b) are the steady state values obtained for the average value of the force $\overline{f}(v)$,
for Glauber and Kawasaki dynamics respectively, as a function of the velocity $v$ for the one dimensional Ising model with parameters, $\beta=1$, $J=1$ and $h=10$.  
Similar results are found in two dimensions. All numerical results have similar characteristics,
the force is always in the opposite direction to the direction of movement, so it is a friction, 
the force is linear for small $v$ and for large $v$ the force decays as $1/v$.  

Fig. (\ref{fig1}) is a plot of the average magnetization around the inclusion seen from its rest frame in the steady state regime with Glauber dynamics. For zero velocity we see that the magnetization profile is spherically symmetric and there is no net force, but as the velocity is increased the magnetization profile is distorted and takes a different form ahead of and behind the inclusion. As $v$  increases the smearing out of the magnetization profile is  increased, but the amplitude of the deformation diminishes.  These results indicate that it is the magnitude of the deformation of the local field by the inclusion along with  the asymmetry induced by the particle motion that induces the force. Once the velocity becomes too large the system does not have time to react to the presence of the inclusion  and 
the frictional force  weakens, this shear thinning like cross over should be quite
generic. These results are qualitatively similar to those of \cite{fus2008} for the interaction of 
an MFM tip passing over a magnetic sample, with dipolar interactions between the tip and the sample, and with spin dynamics described by the  Landau--Lifschitz--Gilbert equation. 

To understand these results  we consider the
general over-damped dynamics for a scalar field $\phi$.  We denote positions in space by ${\bf r} = ({\bf x},z)$ where the coordinate $z$ is in the direction of the motion. We write the Hamiltonian  density $\cal{H}$ as two  parts, the first a bulk term  ${\cal H}_0[\phi]$ and the second the energy due to the inclusion $\delta({\bf x})\delta(z-vt){\cal{H}}_{tr}[\phi]$. We thus assume that the interaction between the inclusion and the field is  localized about the point $(0,vt)$. The energy of the system is thus 
\begin{equation}
H = \int \left({\cal H}_0[\phi] +\delta({\bf x})\delta(z-Z){\cal{H}}_{tr}[\phi]\right)d{\bf r} ,
\end{equation} 
where $Z=vt$ is the inclusion's position in the direction $z$. The instantaneous force acting on the inclusion is simply given by
\begin{equation}
f=-{\partial H\over \partial Z} = -{\partial {\cal{H}}_{tr}(\phi(0,Z))\over \partial Z},
\end{equation}
and when the  derivative is not continuous we implicitly take its average value about the particle
position (corresponding to the symmetric discrete derivative used in the Ising model simulations).
We now take a general over-damped dynamics for the evolution of the field $\phi$
\begin{equation}
{\partial \phi({\bf r})\over \partial t}= -R{\delta H\over \delta\phi({\bf r})} + \eta({\bf r},t),
\end{equation}
where $R$ is a dynamical operator defining the underlying over-damped dynamics. For instance
$R({\bf r},{\bf r'}) =R({\bf r}-{\bf r'})=\delta({\bf r}-{\bf r'})$ for model A dynamics (non-conserved order
parameter) and $R({\bf r}-{\bf r'})=-\nabla^2\delta({\bf r}-{\bf r'})$ for model B dynamics (conserved order
parameter). The noise term is taken to be Gaussian white noise in time  and with spatial correlations
that respect detailed balance, i.e. $\langle \eta({\bf r},t) \eta({\bf r}',t')\rangle = 2T\delta(t-t')R({\bf r}-{\bf r}')$.

Consider the case of free scalar fields with a linear coupling between the inclusion and the
field i.e.
\begin{equation}
{\cal H}_0 = {1\over 2} \phi({\bf r})\Delta\phi({\bf r}),\ \ \ {\cal H}_{tr}= -K\phi({\bf r}),
\end{equation}
where both $\Delta$ and $K$ are linear operators which, for the purposes and applications here,  we will assume to be self adjoint. 
Working in the coordinate system comoving with the inclusion, the evolution equation for average 
value of the field takes the form
\begin{equation}
{\partial {\overline \phi}({\bf r})\over \partial t} -v{\partial {\overline \phi}({\bf r})\over \partial z} = -R\Delta{\overline\phi}({\bf r}) + RK(\bf r) .
\end{equation}
In the steady state we can neglect the time derivative in the above and solve the resulting equation by 
Fourier transforming \cite{comm} (where ${\tilde g}({\bf k}) =\int d{\bf x} \exp(-i{\bf k}\cdot{\bf x}) g({\bf x})$) to obtain 
\begin{equation}
{\tilde{\overline\phi}}({\bf k}) = {{\tilde R}({\bf k}){\tilde K}({\bf k})\over
{\tilde \Delta}({\bf k}){\tilde R}({\bf k})-ik_zv}.\label{mf}
\end{equation} 
A key point of Eq. (\ref{mf}) is that when $v\neq 0$ the average magnetization depends on the 
dynamical operator $R$, it is only when $v=0$ that $\overline{\phi}$ takes its equilibrium  value.
Using this we find the average force to be
\begin{equation}
{\overline f}(v) =  {i\over (2\pi)^{d}}\int d{\bf k}\, {k_z{\tilde R}({\bf k}){\tilde K}^2({\bf k})\over {\tilde \Delta}({\bf k}){\tilde R}({\bf k})-ik_z v},\label{gen}
\end{equation}
where $d$ denotes the spatial dimension. The force thus depends on the particular form of the dynamics
via $R$ as well as the static quantities $\Delta$ and $K$. 
For small velocities we can write this result in terms of an effective friction coefficient ${\overline f}=-\lambda v$ where
\begin{equation}
\lambda= {1\over (2\pi)^dd}\int d{\bf k}{{ k}^2{\tilde K}^2({ k})\over {\tilde \Delta}^2({ k}) {\tilde R}({ k})}\label{fric}
\end{equation}
with $k=|{\bf k}|$ and we have assumed that the operators $\Delta$, $K$, and $R$ are rotationally as well as translationally invariant. For large velocities we find that
\begin{equation}
{\overline f}(v) =  -{1\over v}\times{1\over (2\pi)^{d}}\int d{\bf k}\, {\tilde R}({k}){\tilde K}^2(k)\label{bigv}.
\end{equation}
The limiting forms for the average force for small and large velocities thus take a  universal form
as long as the integrals in Eqs. (\ref{fric}) and (\ref{bigv}) exist. 
These integrals can be regularized at large $k$ by an ultraviolet cut off. However if the insertion has a finite size, then we can approximate  this by replacing the delta function in the above computation
by a Gaussian profile of width $a$ corresponding to the insertion size. This gives $\tilde K(k,a)= \tilde K(k,0) \exp\left(-{k^2a^2\over 2}\right)$,  but this simply sets another cut-off $k_{\rm max} \sim 2\pi/a$.

To see what the generic form of the induced drag is we introduce a time  scale $\tau$ for the dynamics writing, on dimensional grounds, $\tilde R\tilde \Delta= a_0^{\rho+\delta}k^{\rho+\delta}/\tau$, and where for simplicity (however see the discussion later) we assume that $a_0=a$ is the  cutoff of the field  theory as well as the inclusion size. We also write $\tilde \Delta= \mu  a_0^{\delta-d}k^\delta$ and 
 $\tilde K= \mu'  a_0^{\alpha}k^\alpha$. A critical dimension $d_c=2\delta+\rho-2\alpha-2$ is found such that if  $d>d_c$, and we assume that the friction is dominated by the ultra-violet divergence, we  find $\lambda\sim {\mu'^2\over \mu}\tau/a_0^2$. However if $d<d_c$, but there is a mass term $m$ which regularizes the theory at the infrared level then we find $\lambda\sim {\mu'^2\over \mu} \tau (\xi/a_0)^{d_c-d}/a_0^2$, where $\xi=1/m$ is the correlation length of the
field. Hence  when $d<d_c$ the typical friction can be expected to be orders of magnitude
greater than that for systems such that $d>d_c$ if the correlation length is significantly larger than the 
inclusion size. When $d<d_c$, but the system is critical i.e. $m=0$ (corresponding to the Gaussian approximation for continuous phase transitions such as the para-ferromagnetic transition or critical demixing in lipid bilayers) the integral is infrared regularized at $k_{min}\sim 2\pi/L$ where $L$ is the linear dimension of the system and we find $\lambda\sim {\mu'^2\over \mu} \tau (L/a_0)^{d_c-d}/a_0^2$. 

Note that this critical behavior can even be present when the free field theory is {\em not} critical 
($m\neq0$); in this case $d_c = \rho-2\alpha-2$. This strange situation can occur when the long distance modes of the theory are weakly damped ($\rho$ large for $k\ll1$) or when the field coupling to the inclusion position is long range (non local $\alpha <0$). Critical behavior can also be seen in the expression Eq. (\ref{gen}) at small $v$. If $\delta +\rho >1$, which is the case for the examples given in this letter,  and $d<d_c$ we find that ${\overline f}(v) \sim v^{1+ {d-d_c\over \rho+\delta-1}}$.

{\em Gaussian Approximation to the Ising Ferromagnet:} to understand the numerical results on the Ising model, we take, in the disordered phase, the Gaussian model ${\tilde \Delta} = k^2 + m^2$. The coupling to the inclusion is linear so $\tilde{K} = h$. For Glauber dynamics we take the model A dynamics $\tilde{R} = 1$ and for Kawasaki dynamics we take model B with $\tilde{R}=k^2$. When $m=0$, $d_c=2$ for model A and $d_c=4$ for model B.  Shown in Figs (\ref{fig1}a) and (\ref{fig1}b) are the simulation results for the 1d Ising model with Glauber and Kawasaki dynamics fitted with the formula Eq. (\ref{gen}) using model A and B dynamics for the Gaussian model, we see that the fits are excellent in both cases.
  
{\em Drag in lipid membranes:} Saffmann and Delbr\"uck (SD) \cite{saff1975} computed the diffusion coefficient for cylindrical inclusions in membranes   where  the membrane is treated as a continuous medium. This was done by computing the drag  by taking into account the coupling of the 2d hydrodynamic field of the membrane to that of the external medium using the nonlinear Navier Stokes equations.  The diffusion constant is then obtained via the Stokes Einstein relation and is found to depend on the protein radius $a$ as $D\sim \ln(a)$. However a recent experimental investigation  \cite{gam2006} found that the diffusion constant behaves as $D\sim 1/a$. Several factors can change the dependence of the diffusion constant on $a$, for instance with the same hydrodynamical model frictional coupling of the membrane to a solid surface leads to a behavior $D\sim 1/a^2$ \cite{wall}. Deviations from the SD formula have also been attributed  to modifications of membrane thickness, due to hydrophobic mismatch, \cite{naji2007} and also the modification of membrane curvature \cite{naj2009} caused by the protein.

In the spirit of these latter approaches, first consider the  case where the parameter $\phi$ corresponds to the height of the membrane. In the Monge gauge, for small fluctuations, ${\tilde \Delta}(k)  = \kappa k^4 +\sigma k^2$  where $\kappa$ is the bending rigidity and $\sigma$ is the membrane tension \cite{hel1973}. A simple model for the coupling between a protein and a membrane is  a linear coupling between the membrane curvature and the protein (a protein favors a positive of negative curvature of the membrane), this means that $\tilde K = h k^2$. The dynamics of  membrane height fluctuations are dominated by hydrodynamic interactions which  can be shown to lead to a nonlocal dynamical operator $\tilde{R} =  1/4\eta k$, where $\eta$ is the viscosity of the surrounding medium \cite{lin2004}.

To model drag  in membranes we need to know how the magnitude of the coupling $K$ depends on the size of the insertion $a$. To find the dependence of $K$ on $a$ we assume that the energy induced by the inclusion has the form of a line tension for small $a$, that is we impose
\begin{equation}
\epsilon = -{1\over 2 (2\pi)^d} \int d{\bf k} {\tilde K^2(k)\over \Delta(k)} = 2\pi a \gamma,
\end{equation}
where $\gamma$ is the line tension.
We  set $\tilde K = k^\alpha h$ and assume that the integral is dominated either  
by an ultra-violet divergence, or when this is not present by an infrared divergence regulated 
by $k_{min}= m=1/\xi$, yielding,
\begin{eqnarray}
h^2&\sim& a^{d+2\alpha-\delta+1} \ \ \ \ \;{\rm for}\ d+2\alpha -\delta > 0 \\  
h^2&\sim& \xi^{d+2\alpha-\delta}a \; \ \ \ \;\ \ {\rm for}\ d+2\alpha -\delta  < 0 \\  
h^2&\sim& a/\ln\left({\xi\over a}\right) \ \ \ \ \ {\rm for}\ d+2\alpha -\delta = 0 . 
\end{eqnarray}
Using this result to compute the friction for curvature coupling to membrane height fluctuations we find
 $\lambda \sim a^2$. This means that the frictional force is relatively weak, but if it were 
dominant one would estimate, via  the Stokes Einstein relation, that the diffusion constant would scale
as $D=k_BT/\lambda \sim 1/a^2$.  Protein coupling to local lipid composition (either chemical or order) 
can  induce interactions between proteins \cite{sac1995}. Let us take $\phi$ to be either 
the local order parameter specifying the local phase (solid, liquid, gel) for single lipid membranes
or the local lipid composition for membranes composed of lipid mixtures \cite{comm2}, and assume it couples linearly to the protein position, i.e.  $\tilde{K}=h$ is constant.
The experiments  of \cite{gam2006} on the diffusion constant of small peptides were carried out in lipid membranes composed  essentially of a single lipid type SOPC, the order parameter coupled to the protein has no reason to be conserved and so 
we assume dynamics of model A type.   Applying our computations for the magnitude of $h$ along with the dynamical result for the friction Eq. (\ref{fric}) we find  $\lambda \sim a$. Thus for small insertions this drag force is expected to be larger that the height fluctuation induced force and if this drag dominates we obtain the behavior $D\sim 1/a$ --  the scaling found in \cite{gam2006}. It is thus possible that, for small membrane inclusions,  the dominant drag force is due to interaction with an order parameter of the  surrounding lipids.  

We have shown how inclusions moving at constant velocity in classical fields experience a
drag force. This drag force is generated by a polaron-like perturbation of the surrounding field, at large
velocity the polaron is weakened as the field cannot react sufficiently quickly to the inclusion. The results 
given here are for free fields, but a rich phenomenology arises from a relatively straightforward analysis. The wealth of the theory comes in part from the range of field theories that
describe soft-matter systems and from the range of dynamical models that are present. There are clearly
a wide range of directions for further study, including (i) the effect of interacting fields, (ii) extensions
to more realistic dynamics and (iii) the possibility of observing the drag behavior predicted here in experiments. For example this phenomena could possibly  be studied using optical tweezers to pull an inclusion, such as a colloid, through a binary fluid mixture, in particular at its critical point.  It would also
be interesting to see how protein coupling to local membrane composition affects the diffusion constant
for proteins in multicomponent membranes.


\begin{thebibliography}{99}
\bibitem{comm}{Solving the problem in Fourier space ensures that derivatives so computed
are averaged over their left and right values (Dirichlet's theorem).}
\end{thebibliography}
\end{document}